\documentclass[prl,preprint,showpacs,superscriptaddress,floatfix]{revtex4}
\usepackage[latin1]{inputenc}
\usepackage{graphicx}
\usepackage{float}
\usepackage{amsmath}
\usepackage{natbib}

\begin{document}
\title{Extraordinary absorption of sound in porous lamella-crystals}

\author{J. Christensen*}
\affiliation{Institute of Technology and Innovation,
University of Southern Denmark, DK-5230 Odense, Denmark}
\affiliation{Department of Photonics Engineering, Technical University of Denmark, DK-2800 Kgs. Lyngby, Denmark}

\author{V. Romero-Garc\'ia}
 \affiliation{Instituto de Investigaci\'on para la Gesti\'on Integrada de zonas Costeras, Universitat Polit\`ecnica de Val\`encia, Paranimf 1, 46730, Gandia, Spain}
 \affiliation{LUNAM Universit\'e, Universit\'e du Maine, CNRS, LAUM UMR 6613, Av. O. Messiaen, 72085 Le Mans, France}
 
\author{R. Pic\'o}
 \affiliation{Instituto de Investigaci\'on para la Gesti\'on Integrada de zonas Costeras, Universitat Polit\`ecnica de Val\`encia, Paranimf 1, 46730, Gandia, Spain}
 
 \author{A. Cebrecos}
 \affiliation{Instituto de Investigaci\'on para la Gesti\'on Integrada de zonas Costeras, Universitat Polit\`ecnica de Val\`encia, Paranimf 1, 46730, Gandia, Spain} 

 \author{F.~J.~Garc\'ia~de~Abajo}
 \affiliation{ICFO Institut de Ci\`encies Fot\`oniques, Mediterranean Technology Park, 08860 Castelldefels (Barcelona), Spain}  

 \author{N. A. Mortensen}
 \affiliation{Department of Photonics Engineering, Technical University of Denmark, DK-2800 Kgs. Lyngby, Denmark}

 \author{M. Willatzen}
 \affiliation{Department of Photonics Engineering, Technical University of Denmark, DK-2800 Kgs. Lyngby, Denmark}
 
 \author{V. J. S\'anchez-Morcillo}
 \affiliation{Instituto de Investigaci\'on para la Gesti\'on Integrada de zonas Costeras, Universitat Polit\`ecnica de Val\`encia, Paranimf 1, 46730, Gandia, Spain} 
 
\date{\today}

\begin{abstract}

We present the design of a structured material supporting complete absorption of sound with a broadband response and functional for any direction of incident radiation. The structure which is fabricated out of porous lamellas is arranged into a low-density crystal and backed by a reflecting support. Experimental measurements show that strong all-angle sound absorption with almost zero reflectance takes place for a frequency range exceeding two octaves. We demonstrate that lowering the crystal filling fraction increases the wave interaction time and is responsible for the enhancement of intrinsic material dissipation, making the system more absorptive with less material. 
\end{abstract}
\date{\today}
\maketitle

The damping of sound waves can be understood by conversion of the mechanical energy into heat. The acoustic equivalent to an ideal black body would be something similar to a "deaf" body which is an object absorbing sound coming from all directions at any given frequency. Absorption of sound waves is governed by the effects of viscosity and thermal conduction in fluids. In order to give a macroscopic picture of these dissipative processes one must interchange the mass density $\rho$ and the bulk modulus $K$ with complex quantities, leading to a finite penetration length into the dissipative medium. In a first approximation, damping can be neglected for audible sound in free space and therefore the utilization of lossy materials is essential for screening noise and designing acoustic insulators. Thus, it is desirable to create structures with the capability of efficient dissipation, preferably in a way that energy conversion causes all waves to be absorbed such that no back- and through-radiation takes place. While this remains to be the ideal case, various composites and artificial structures were designed in the attempt of pursuing this ultimate goal. We distinguish between resonance-based and broadband systems. Locally resonating materials have been fabricated in the form of mass-loaded thin membranes, gas-bubble arrays and elastic beams, featuring sharp and narrow absorption peaks \cite{DarkPing,JohnPageEuro,Finkbubble,Naturebubble,Romero}. Broadband absorption, on the other hand, has been demonstrated for low frequencies by lattices of perforated shells \cite{Dehesa}. Sound blocking screens made out of, e.g., phononic crystal or metamaterials with one single negative effective parameter prevent waves to penetrate through these structures resulting in full reflection of sound \cite{Kushwaha,Vasseur,Zhengyou,allAngleBlocke}. These effects are caused by diffraction properties and evanescent modes, respectively, and will not easily lead to broadband absorption of sound. Periodic penetrable structures, generally speaking, have been fabricated with many facets for different kinds of waves. Electromagnetic (EM) structured materials, in that regard, such as gratings with finite conductance, convex grooves or the moth eyes, comprise anti-reflective systems with a broad spectral response \cite{McPhedran1,McPhedran2,PlasmonicAbs,BlackGold,MothEye}. However, to sustain a spectrally and angularly rich performance with complete absorption and little material use remains a challenge worth pursuing.\\
We present a system inspired by recent EM experiments where a forest of vertically aligned single-walled carbon nanotubes showed extremely low reflectance and making it the darkest man-made material ever \cite{VidalEffective,ExpObserv,NviewsVidal,PNASblack,ACSblack}. The mechanism consists in the attempt of matching the material index to free-space to prevent back-reflected waves but at the same time providing sufficient material losses to guarantee intensity attenuation. Here, we show how a low-density porous lamella array, in analogy to its electromagnetic counterpart made of nanotube arrays, behaves most closely like a true deaf body. Within this framework, we show how these constructed crystals in a most counter intuitive way become more absorptive when less material is chosen.\\
To illustrate this surprising behaviour, we begin by analysing a simple asymmetric system consisting of an inhomogeneous lossy medium of length $L$ supported by a rigid backing as illustrated in Fig. 1a. The general complex scattering matrix can be written as:
\begin{equation}
\label{1st}
{S}(\omega)=
    \begin{bmatrix}
      r(\omega) & 0 \\[0.3em]
      0 & 1 \\[0.3em]
     \end{bmatrix}, 
\end{equation}
where $r(\omega)$ is the complex reflection coefficient. In the absence of losses the scattering matrix is unitary and consequently $|r(\omega)|$ = 1, i.e. perfect reflection due to the rigid backing. This changes dramatically in the presence of weak absorption provided that the time delay is comparable to $1/2\Gamma$. To see this explicitly, we follow considerations developed in the context of absorption in chaotic cavities  \cite{Chaos}. For a weak absorption rate $\Gamma$ the full scattering matrix $S$ is related to its loss-less counterpart $S_0$
\begin{equation}
\label{2nd}
{S}(\omega)=S_0(\omega)\left(1-\frac{\Gamma}{2}Q(\omega)\right), 
\end{equation}
where $Q(\omega)=-iS_0^\dagger(\omega)\frac{\partial S_0(\omega)}{\partial \omega}$ is the Wigner-Smith delay-time matrix. Since $S_0$ is Hermitian so is $Q$ and its positive eigenvalues are interpreted as the group delay time $\tau$ for a wavepacket centered around $\omega$. Conservation of energy in the asymmetric system requires $|r(\omega)|^2+A=1$, where $A$ is the absorption, hence, to a first approximation we can write 
\begin{equation}
\label{3rd}
|r(\omega)|^2=1-2\Gamma\tau. 
\end{equation}
Provided a sufficiently long time delay, the reflectance can be minimal and the absorption perfect, even for a modest intrinsic absorption rate. The perspective of acoustic functional materials is to support delay times much exceeding the delay time of their bulk counterparts. For a bulk material lamella ($l$) of length $L$, backed by a perfect reflector, the acoustical path becomes $2L$ and consequently the time delay is $\tau_l=2L/c_l$ where $c_l$ is the sound speed in the lamella material. On the other hand, composing a porous periodic crystal from the same base material, $c_l$ is replaced by the actual group velocity $v_g$ associated with the phononic band dispersion relation. In this context, we define the enhancement factor $\gamma=\frac{\Gamma\tau}{\Gamma_l\tau_l}$ that expresses the acoustic interaction strength in the crystal. In other words, enhancement is reached whenever $\gamma>1$, meaning that dissipation inside the crystal exceeds the intrinsic material losses of the lamellas. From Eq.~(\ref{3rd}) it follows that reflectance (absorption) will decrease (increase) whenever wave slowing or increased dissipation take place. Since we will not operate within the long wavelength regime, we apply this model in our interpretation of the enhanced absorption because homogenization of the lamella-crystal cannot be applied safely.\\
\linebreak We examine the structure depicted in Fig. 1a where the goal is to explore the tunability for maximal absorption. The basic structure is a 1D sonic crystal consisting of an array of lossy slices supported by a perfectly rigid backing. With translational invariance in the $y$ direction we numerically simulate the complex wave interaction by coupling free-space sound radiation to Bloch states inside the crystal (detailed derivations are found within the Supplementary Informations \cite{Suppl}).
\begin{figure}[]
\begin{center}
\vspace{-1.5cm}
\includegraphics[width=0.6\columnwidth]{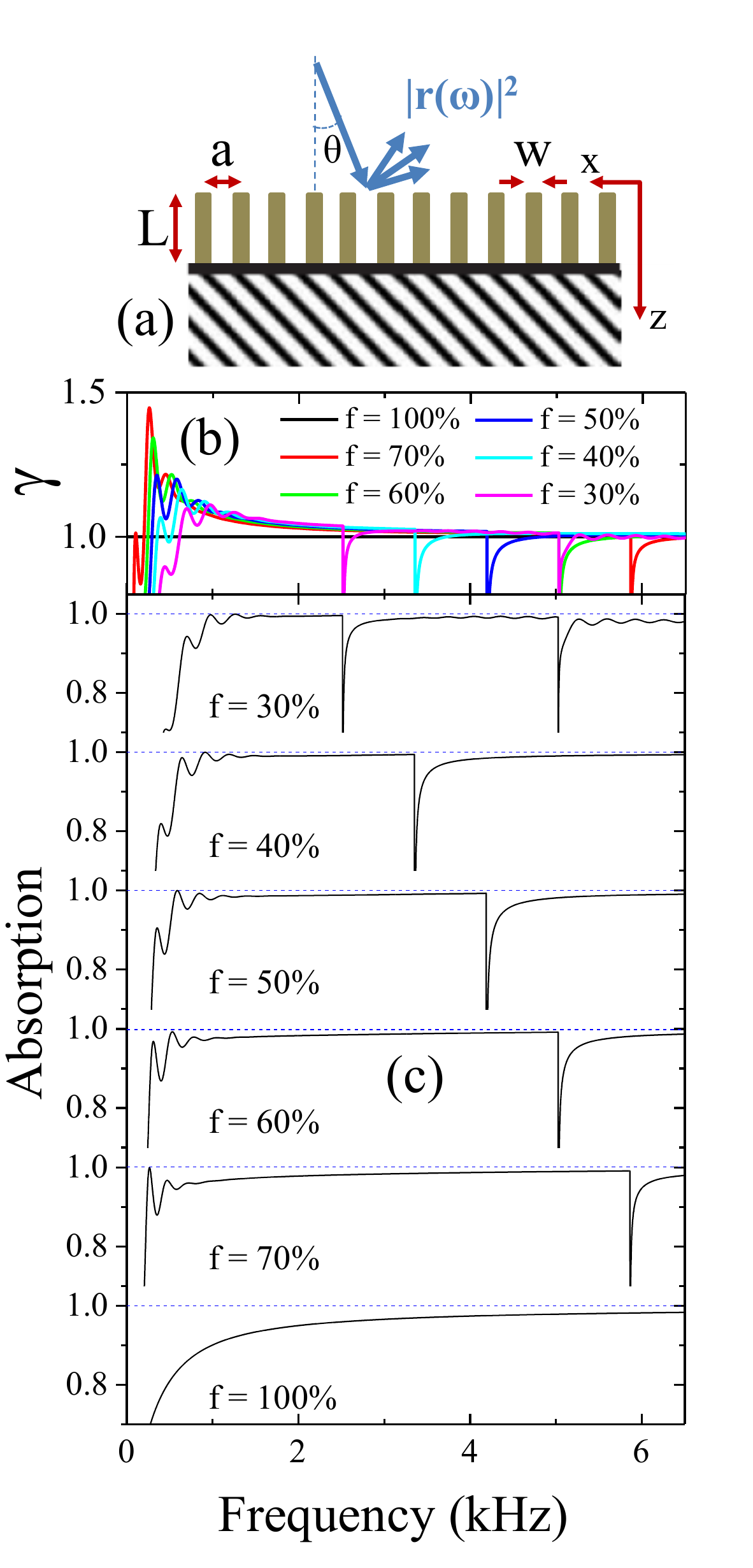}
\caption{Controlling the absorption by lowering the crystal filling fraction. \textbf{a}, Schematic of the crystal made of lamellas of width $w$, lattice constant $a$ and length $L$. The crystal is backed by a rigid support into which no sound waves penetrate. \textbf{a,b}, The enhancement factor $\gamma$ and the absorption are plotted for various filling fractions with constant lamella width, $w=4$ cm and length $L=0.5$ m. In all cases, we implemented complex dispersive material dependence for $\rho$ and $K$ within the lamellas \cite{Lafarge97}.}
\label{fig:fig1}
\end{center}
\end{figure}
This leads to the following expression for the reflection coefficient
\begin{equation}
\label{4th}
\begin{aligned}
r_\textbf{G}=2\sum_{j}p_j^\textbf{G}B_j\text{sin}\;q_jL-\delta_{0\textbf{G}}, 
\end{aligned}
\end{equation}
where $p_j^{\textbf{G}}$, $B_j$ and $q_j$ are the $j$'th eigenvector, modal amplitude and out-of-plane wave vector of the crystal respectively, from which we derive the overall absorption
\begin{equation}
\label{5th}
\begin{aligned}
A=1-\sum_{\textbf{G}}\text{Re}\left(\frac{k_z}{k_z^\textbf{G}}\right)|r_\textbf{G}|^2, 
\end{aligned}
\end{equation}
where $k_z^\textbf{G}=\sqrt{k_0^2-\left(k_x+G_x\right)^2}$. We begin by analysing the spectral absorption $A$ ( Eq. (\ref{5th})) for sound incident along the out-of-plane crystal orientation at normal incidence. In the experimental set-up we utilize a porous material based on homogeneous foam, hence to conduct numerical simulations we have implemented dispersive material dependence for these lossy materials and calculated the acoustical response of the crystal to an incoming sound wave \cite{Lafarge97}. In order to lower the filling fraction and by this controlling the overall dissipation, we need to change the lattice constant for a fixed lamella width. By inspecting the bulk properties for the case when $f=100\%$ or equivalently $\gamma=1$ we predict overall weak absorption for low frequencies as seen in Fig. 1c. Obviously this is the regime to be challenged since porous materials naturally perform most efficient at higher frequencies. Changing the crystal filling fraction $f$ by varying the lattice parameter $a$ causes the lattice singularities to shift accordingly as seen in Fig. 1c. These spectral dips resulting in low absorptions arise from diffracted waves becoming grazing and take place when $|k_x+G_x|=\omega/c_0$, with $\textbf{G}=G_x$ \cite{deAbajo2007}. If the filling fraction is lowered a growth in the enhancement factor is seen, which is giving rise to increased absorption. This increase in dissipation is simply explained by the fact that sound is trapped more efficiently inside the crystal as compared to the bulk ($f=100\%$) meaning that $\gamma$ must be larger than unity as depicted in Fig. 1b. We obtain enhanced absorption caused by an increased acoustic interaction strength ($\gamma>1$) that for higher frequencies slightly grows inversely with the filling fraction, see Fig. 1b.
\begin{figure}[htbp]
\begin{center}
\vspace{-0.5cm}
\includegraphics[width=0.6\columnwidth]{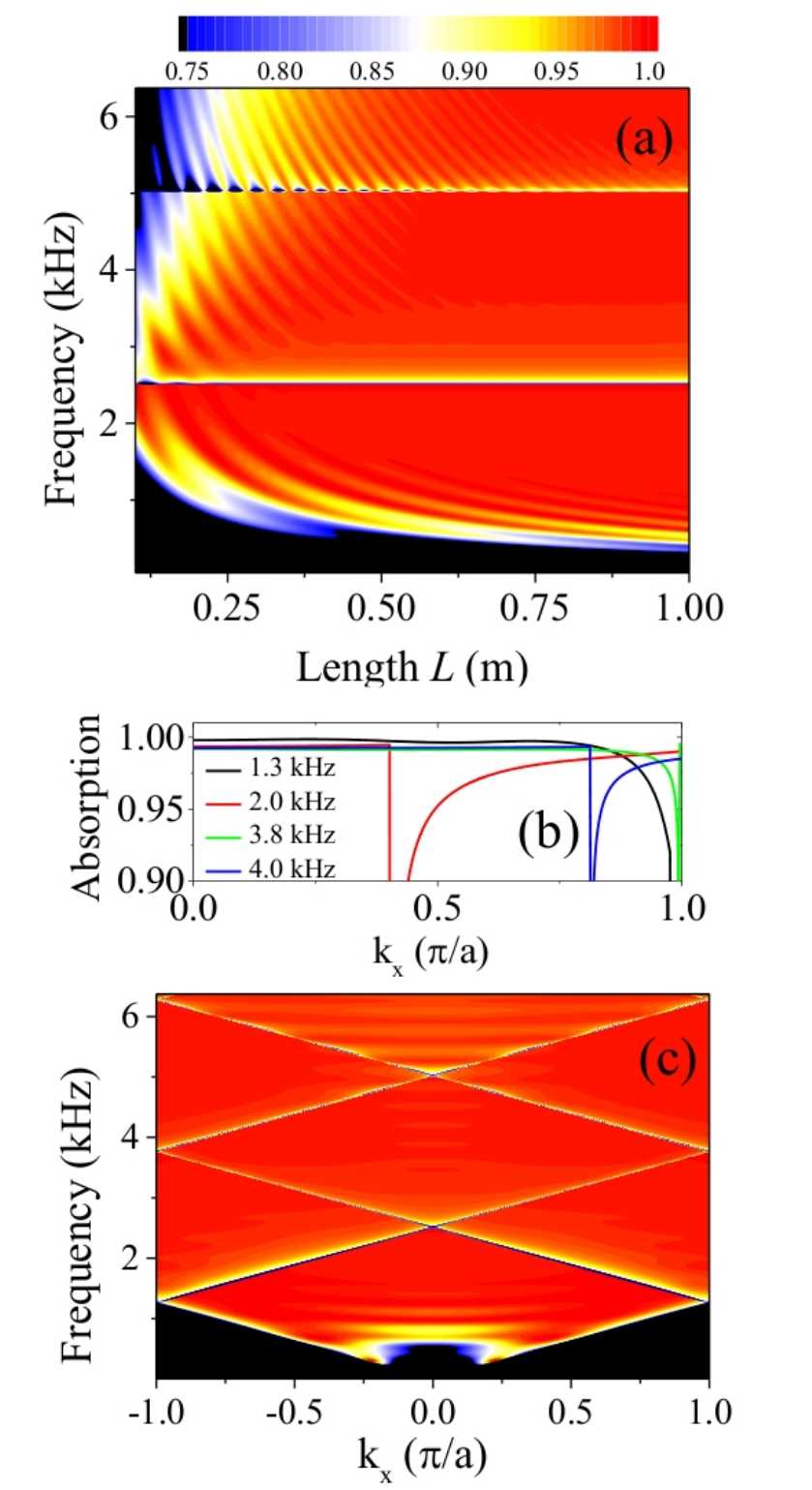}
\caption{Acoustic response of crystals made out of porous lamellas. \textbf{a}, Absorption versus frequency and lamella length $L$ for normal incident radiation where $f=30\%$ and $w=4$ cm. \textbf{b,c} Simulated angular dependence of the structure in \textbf{a} with $L=0.5$ m within the first Brillouin zone. We take illustrative examples corresponding to four frequency cuts in \textbf{c} and show the dependence of absorption with parallel momentum $k_x$.}
\label{fig:fig2}
\end{center}
\end{figure}
On the other hand, one could in as much create systems with complete absorption $\gamma>1$ of sound by designing lamellas made out of an intrinsic lossier material. It is the choice of materials with high absorption rates and the resulting delay time within the dispersive crystal which dictates the performance. \\
Further to this, we simulate the dependence of the absorption with lamella length $L$. At a given frequency the waves penetrate into the crystal with a characteristic length that should not surpass the crystal length $L>l_p$ to avoid strong back-reflection at the rigid support. For this reason, either the wave needs to decay rapidly or travel a sufficient long distance to undergo enough intensity attenuation as plotted in Fig. 2a, showing that absorption increases when the structure becomes lengthy. In Fig. 2a we also observe some fine oscillations in the mapping of which the spectral locations scale with $1/L$ as known from cavity resonances. More importantly however, although we predict high absorptions in the range 0.95-0.999 in a representable band between $2.0$ to $2.5$ kHz for $L=0.1$ m, as seen in Fig. 2a the bandwidth of complete sound absorption is easily broadened by increasing the lamella length $L$. Unlike membrane type absorbers decorated with rigid platelets where the absorption is resonance-based and overall narrow in width~\cite{DarkPing}, the present scheme relying on diffraction constitutes an acoustically thick layer which is essential for broadband applications. Next, we investigate the angular sensitivity for the crystal, which up to this point has been simulated for normal incident acoustic plane waves only.
\begin{figure}[]
\begin{center}
\centerline{\includegraphics[width=1.2\columnwidth]{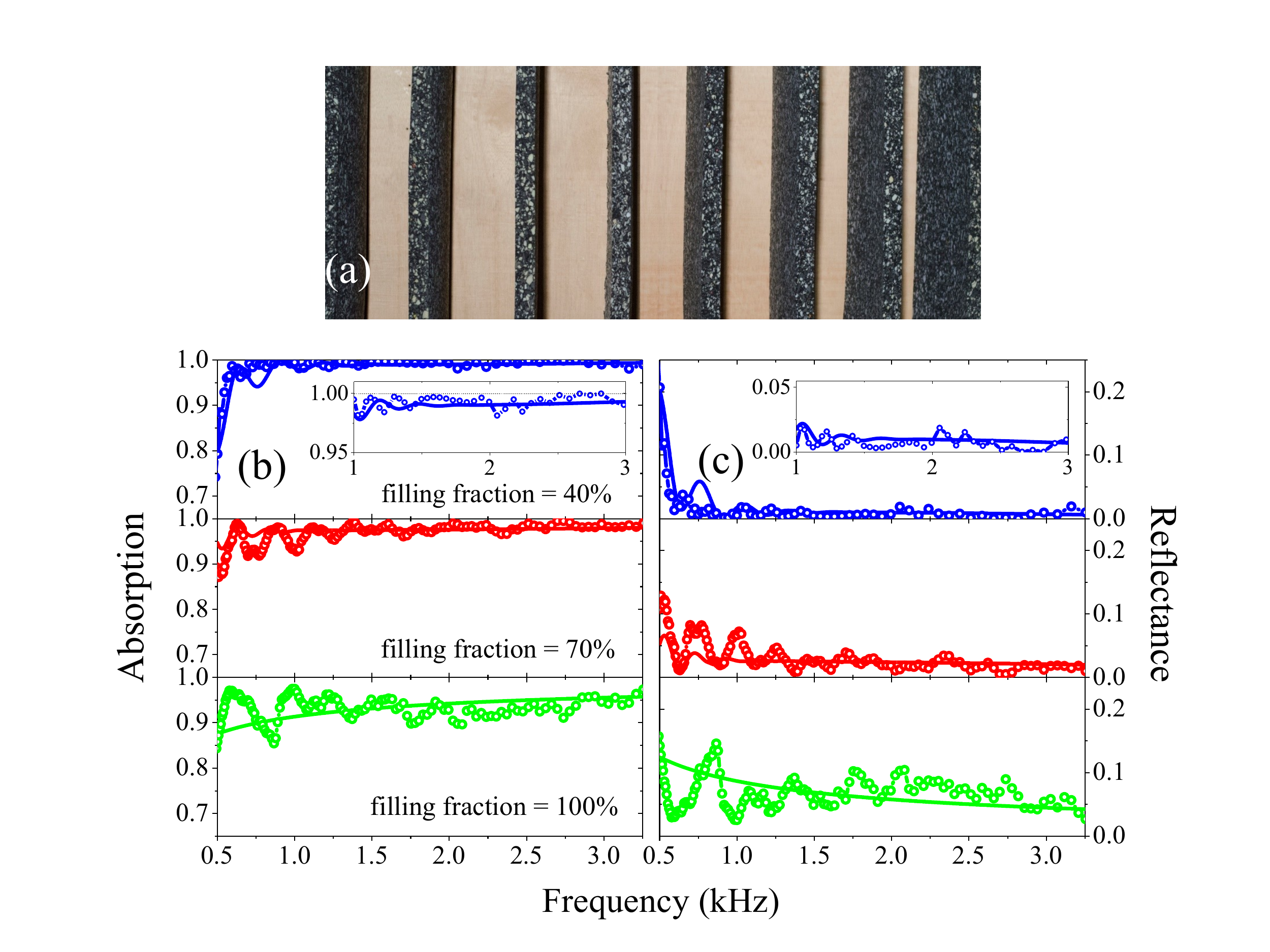}}
\caption{Measured and simulated absorption and reflectance demonstrating increasing dissipation when lowering the filling fraction. \textbf{a}, Bottom view on the suspended crystal made out of porous foam lamellas and supported by a rigid backing. The lamellas have the width $w=4$ cm and length $L=0.5$~m. \textbf{b,c}, Measured (open circles) and simulated (full lines) absorption and reflectance versus frequency for three different filling fractions, $f=40\%$, $70\%$ and $100\%$. All data are obtained for a normal incident acoustic plane wave.}
\label{fig:fig3}
\end{center}
\end{figure}
For this, it is useful to calculate the absorption versus frequency and parallel momentum $k_x$ within the first Brillouin zone. Fig. 2c shows this band diagram containing Bragg-folded sound lines which are responsible for the lattice singularities of low absorption. The entire angular and spectral response resembles the one seen with bandgap materials with no material contrast, but for the present case comprising out-of-plane propagation, it is shown that near to $100\%$ absorption sustains for the angles shown within the first Brillouin zone (see Fig. 2b for specific frequency cuts). Beyond this zone, absorption remains high up to grazing incidence and the limits are met only whenever irradiation is in phase with the crystal lattice and full reflection takes place.\\
 The crystal is constructed out of thin layers of homogeneous foam that is created by compaction and compression of a polyurethane, polyester and polyether mixture. These lamellas have a width of $w=4$ cm and are arranged into a 1D lattice, which is suspended and backed by a rigid wood support as illustrated in Fig. 3a. We have applied the transfer function method consisting of a loudspeaker and two microphones for the phase and amplitude measurements to detect complex reflections. In the experimental set-up the loudspeaker is placed at a sufficient distance from the microphones and the sample to ensure plane wave generation for all relevant frequencies (see Supplementary Informations \cite{Suppl}). Spectrally, we evaluate the reflectance $|r(\omega)|^2$ and the resultant absorption $A$ over frequency ranges relevant to road and air traffic-noise screening. From the previous study we predicted that lowering the crystal filling fraction will improve the absorption of sound due to the enhancement of the interaction strength. To validate this experimentally we have constructed various samples made out of the same lamellas but varied the size of the unit cell resulting in three different filling fraction as seen in Fig. 3. Bulk material properties are obtained measuring the response of the structure with a filling fraction of 100\% from which relatively strong absorption stems from intrinsic material losses and the slab length $L$. When lowering the filling fraction down to 70\% and 40\%, we observe increased performances with mean absorption of 0.97 and 0.99, respectively, over an extended spectral range spanning from $0.7$ to $3$kHz. Due to a finite number of unit cells (\cite{Suppl}) we detect oscillations in the spectrum, overall however, the theory agrees very well with the average absorption evaluated from experiments. 
\begin{figure}[]
\vspace{-1.5cm}
\begin{center}
\centerline{\includegraphics[width=0.7\columnwidth]{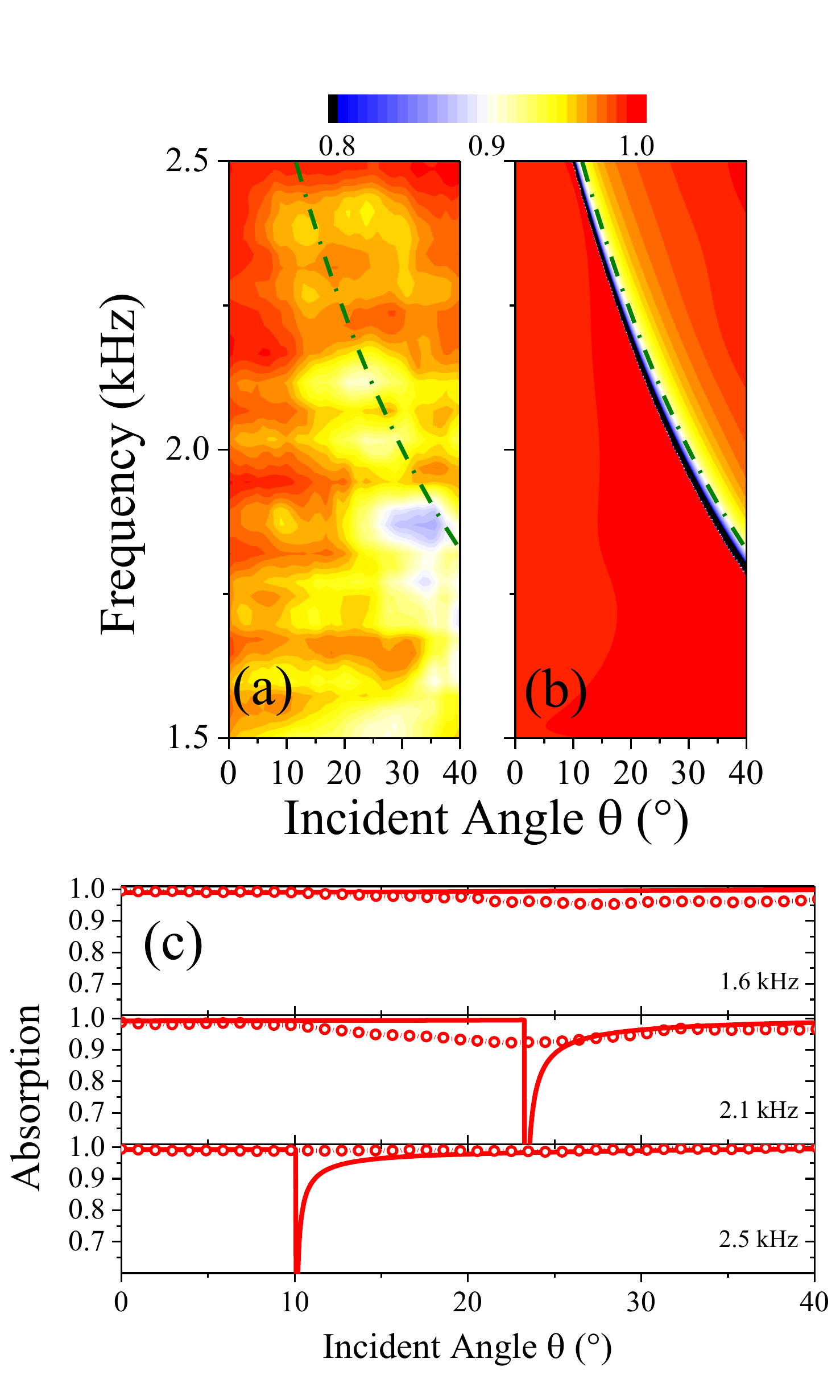}}
\caption{Measured and simulated angle-resolved absorption spectrum. \textbf{a}, Measured (see \cite{Suppl} for details) and \textbf{b}, simulated absorption as a function of frequency and angle of incidence for the same structure as in the previous measurements, now just with $f=36\%$. The dash-dotted line on both contour-maps indicate the condition where $|k_x+G_x|=\omega/c_0$. \textbf{c}, Angular response of the absorption, representing three measured (open circles) and simulated (full lines) frequency cuts from the contours in \textbf{a} and \textbf{b}.}
\label{fig:fig4}
\end{center}
\end{figure} 
The crystal fabricated with a filling fraction of 40\% does not only outperform the other samples measured, see insets of Fig. 3, but further to this we measure a reflectance on the order of $10^{-4}$ in a representable frequency range.\\
Due to the interaction strength exceeding unity, complete sound absorption extends much further away from normal incidence for various directions ($k_x=k_0\text{sin}\theta$) as rendered in Fig.~2c. We conducted angular-resolved absorption measurements where we vary the angle $\theta$ from the normal to the crystal surface. Caused by diffraction associated by momentum transfer to the lattice (dash-dotted lines), again we predict regions of higher reflections as seen in Fig. 4. This narrow region is however being surpassed by a spectrally broad region of strong absorption spanning from $\theta=0^\circ-40^\circ$. Both, the finite number of unit cells and the rigid rotatable frame used for the angular measurements (\cite{Suppl}) are causing additional unwanted reflections in the measurements, Fig. 4a. However, upon inspecting both experimentally and numerically the absorption for three different frequencies within the entire resolved spectra (Fig. 4a and 4b, respectively), we find overall good agreement validating broadband absorption for almost any direction as seen in Fig. 4c. \\ 
We have shown a new concept to engineer high absorptive acoustic materials by means of increasing the sound-material interaction strength which is leading to vanishing reflectance, hence, perfect absorption of sound. We have achieved this goal by fabricating crystals made out of intrinsically lossy lamellas and mounted them onto a rigid backing. For a representable broad range of frequencies we show that absorption can be controlled by the crystal density and tuned up to 99\%. The counterintuitive ability to increase absorption by lowering the effective amount of material is providing many interesting opportunities for minimizing noise pollutions by producing efficient acoustic sealing. \\
The work was supported by the Spanish Ministry of Science and Innovation and European Union FEDER through project FIS2011-29734-C02-01. J. C. gratefully acknowledges financial support from the Danish Council for Independent Research and a Sapere Aude grant (12-134776). V. R. G. gratefully acknowledges financial support from the "Contratos Post-Doctorales Campus Excelencia Internacional" UPV CEI-01-11. J. C. would like to thank Jean-Fran\c{c}ois Allard for stimulating discussion.
*jochri@fotonik.dtu.dk

\end{document}